\documentclass[twoside,12pt]{article}
\usepackage{epsfig}
\usepackage{subfigure}

\newcommand{\be}{\begin{equation}}
\newcommand{\ee}{\end{equation}}
\newcommand{\bea}{\begin{eqnarray}}
\newcommand{\eea}{\end{eqnarray}}

\begin{document}

\title{ \vspace{1cm} Pygmy Dipole Resonance in Exotic Nuclei}
\author{N.\ Tsoneva,$^{1,2}$ H.\ Lenske$^1$ \\
\\
$^{1}$ Institut f\"ur Theoretische Physik, Universit\"at Giessen, 35390 Giessen, Germany\\
$^2${Institute for Nuclear Research and Nuclear Energy, 1784 Sofia, Bulgaria}\\
\\
}
\maketitle
\begin{abstract}
The evolution of the PDR strength with the neutron excess is investigated in Sn isotopic and N=82 isotonic chains with regard to its possible connection with the neutron skin thickness. For this purpose a recently proposed method incorporating both HFB and multi-phonon QPM theory is applied. Analysis of the corresponding neutron and proton dipole transition densities is presented.
\end{abstract}
\section{Theoretical approach and results}
The existence of low-energy dipole excitations located close to the particle emission threshold is observed as a common feature in many isospin asymmetric atomic nuclei. These modes are related to the so-called Pygmy Dipole Resonance (PDR) which has been explained in  neutron-rich nuclei as generated by oscillations of weakly bound neutrons
in respect of isospin symmetric core  \cite{PLB2004,Rez,Paar,Volz}.

In our approach the nuclear ground states are determined by a procedure based on a fully microscopic HFB description  \cite{PLB2004,Hofmann}.
\begin{center}
\begin{figure}[htb]
\includegraphics[width=8.5cm,height=6cm]{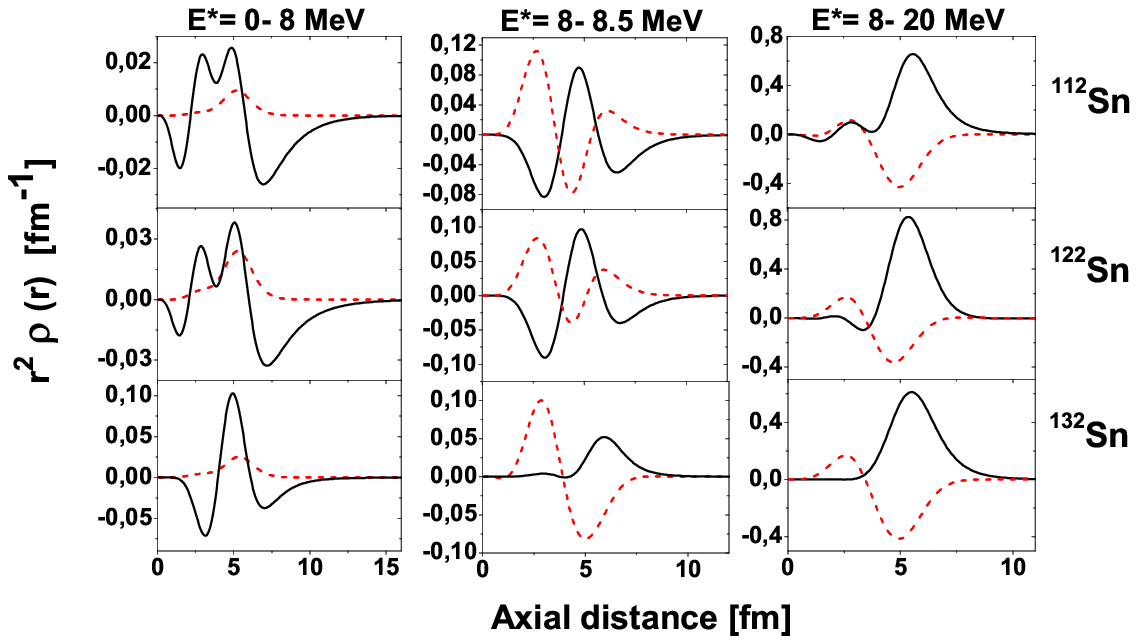}
\includegraphics[width=9.5cm,height=6cm]{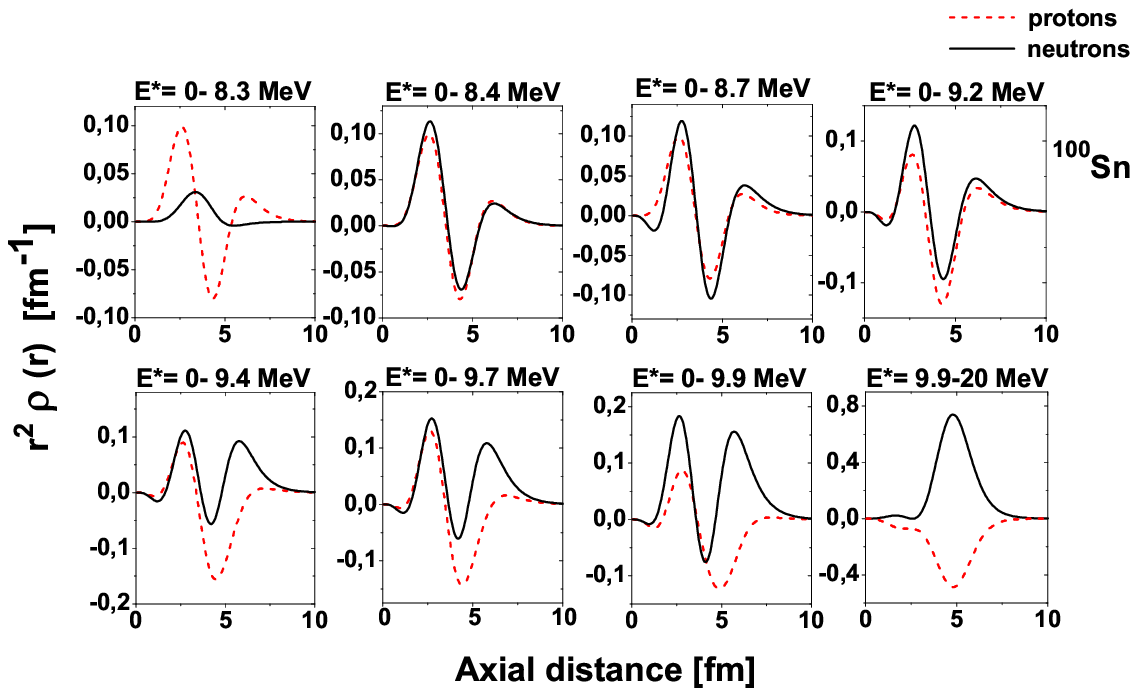}
\caption{\label{fig:fig1} Calculations on one-phonon dipole transition densities in Sn nuclei for different regions of excitation energies E*.}
\label{fig:fig1}
\end{figure}
\begin{figure}[htb]
\includegraphics[width=9cm,height=6cm]{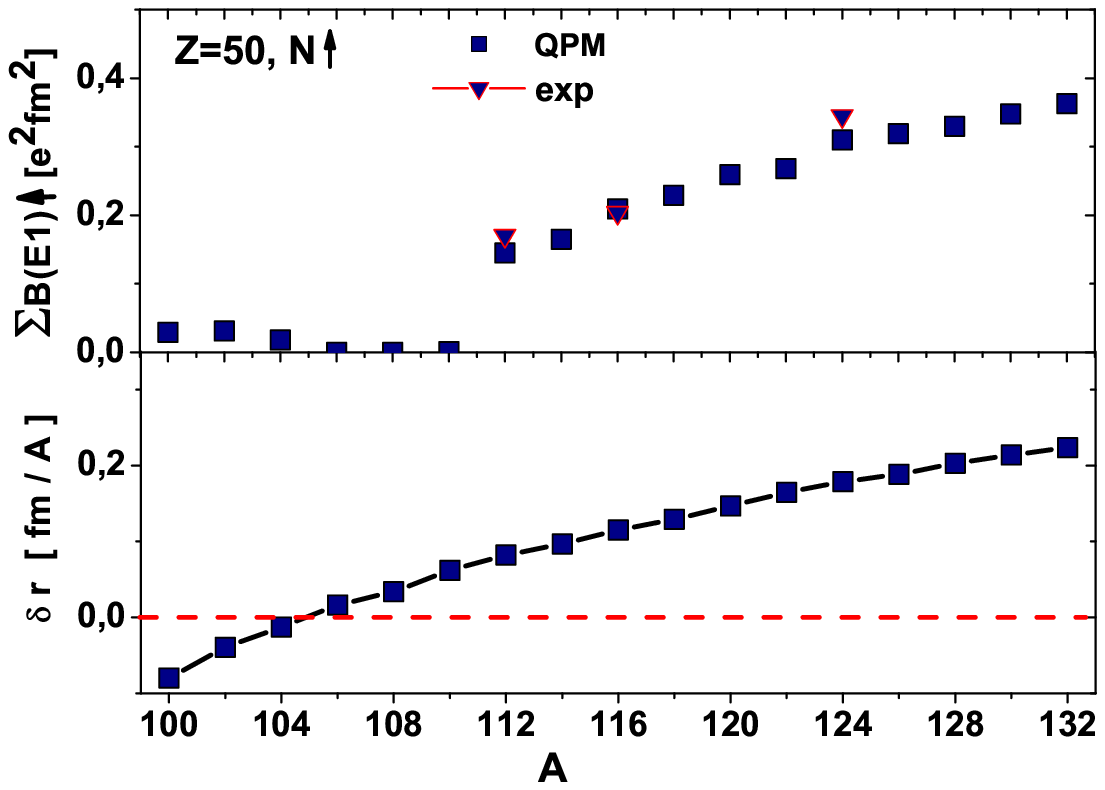}
\includegraphics[width=5cm,height=6cm]{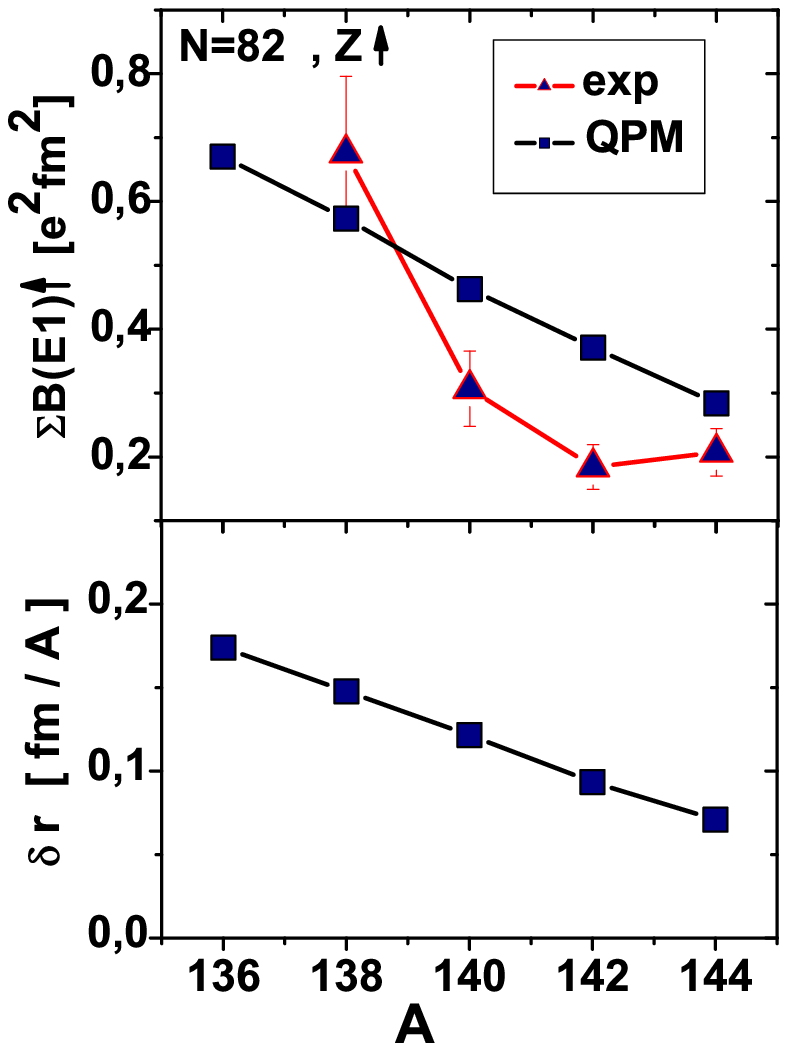}
\caption{\label{fig:fig1} QRPA calculations on the total PDR strength as
a function of the neutron excess in comparison with the relative difference of neutron and proton rms radii $\delta$r in Sn isotopes and N=82 isotones.}
\label{fig:fig1}
\end{figure}
\end{center}
For description of the excited states QPM theory is applied \cite{Sol}. From QRPA calculations in $^{112-132}$Sn and N=82 nuclei a sequence of  low-lying one-phonon dipole states (E*=6-7.5MeV) of almost pure neutron structure is obtained. The analysis of the dipole transition densities at E*$\le$ 8 MeV in $^{112-132}$Sn and N=82 nuclei (Fig. 1) reveal in-phase oscillation of protons and neutrons in the nuclear interior, while at the surface only neutrons contribute. The states in the region E*= 8-8.5 MeV carry a different signature, being compatible with the low-energy part of the GDR. At E*= 8.5-20 MeV a strong isovector oscillation corresponding to the excitation of the GDR is obtained. An interesting observation is the most exotic $^{100}$Sn nucleus where at E*=8.3 MeV a state with a proton structure is found. The detailed analysis on dipole transition densities for different  excitation energy regions in $^{100}$Sn is presented in Fig.1, illustrating the proton surface oscillations at E*$\leq$8.3 MeV. This mode could indicate a proton PDR. The dependance of the calculated total PDR strength on the mass number in $^{100-132}$Sn and N=82 nuclei is presented in comparison to the relative difference between the neutron and proton rms radii $\delta$r in Fig.2. The total PDR strength increases when $\delta$r increases and correspondingly the neutron or proton skin thicknesses increase.

In conclusion, from the analysis of neutron and proton transition densities in  $^{110-132}$Sn isotopes and in N=82 isotones we found almost pure neutron oscillations at the nuclear surface, identified with a PDR. The total PDR strength is closely correlated with the size of the neutron (or proton) skin. The  transition densities act as clear theoretical indicators characterizing the new quality of the PDR mode as a distinct and unique excitation, different from the well known GDR. An interesting result is the observation of proton kind of PDR for the nuclei $^{100-104}$Sn.
The calculations in Sn isotopes and N=82 nuclei are in a good  agreement with available data \cite{Volz,Gov,Ozel}.


\begin{thebibliography}{99}
\itemsep -2pt
\bibitem{PLB2004} N. Tsoneva, H. Lenske, Ch. Stoyanov,
Phys.\ Lett.\ {\bf B586} (2004) 213.
\bibitem{Rez}N. Ryezayeva, T. Hartmann, Y. Kalmykov et al.,
Phys. Rev. Lett. 89 (2002) 272502.
\bibitem{Paar} N. Paar, T. Niksic, D. Vretenar, P. Ring, Phys.~Rev.~{\bf C67}, 034312 (2003).
\bibitem{Volz}S. Volz, N. Tsoneva, M. Babilon et al., Nucl.~Phys.~{\bf A 779} (2006) 1.
\bibitem{Sol} {V.G.~Soloviev, {\it Theory of complex nuclei}
(Oxford: Pergamon Press, 1976).}
\bibitem{Hofmann} F. Hofmann and H. Lenske, Phys.~Rev.~{\bf C57} (1998) 2281.
\bibitem{Gov}K.\ Govaert, F.\ Bauwens, J.\ Bryssinck et al.,
Phys.\ Rev.\ {\bf C57} (1998) 2229.
\bibitem{Ozel} B. $\ddot{O}$zel, P. von Neumann-Cosel, private communications.%
\end{thebibliography}
\end{document}